\documentclass[prd,aps,floats,twocolumn,twoside,preprintnumbers,
superscriptaddress,floatfix,nofootinbib,showpacs]{revtex4}
\usepackage{graphicx,pstricks,rotating,amsmath,epstopdf}
\usepackage{amssymb,bbm}
\usepackage{slashed}
\begin{document}

\title{ Equivalence and/or quantum duality of the non/supersymmetric
noncommutative field theories related by the $\theta$-exact
Seiberg-Witten map$^{\star}$}

\author{Raul Horvat}
\affiliation{Institute Rudjer Bo\v{s}kovi\'{c}, Division of Experimental Physics, Bijeni\v{c}ka 54, 10000 Zagreb, Croatia}
\email{horvat@irb.hr}
\author{Carmelo P. Martin}
\affiliation{Departamento de F\'{\i}sica Te\'orica I, Facultad de Ciencias F\'{\i}sicas,
Universidad Complutense de Madrid, 28040-Madrid, Spain}
\email{carmelop@fis.ucm.es}
\author{Josip Trampeti\'{c}}
\affiliation{Institute Rudjer Bo\v{s}kovi\'{c}, Division of Experimental Physics, Bijeni\v{c}ka 54, 10000 Zagreb, Croatia}
\affiliation{Max-Planck-Institut f\"ur Physik, (Werner-Heisenberg-Institut), F\"ohringer Ring 6, D-80805 M\"unchen, Germany}
\email{josip.trampetic@irb.hr}
\author{Jiangyang You}
\affiliation{Institute Rudjer Bo\v{s}kovi\'{c}, Division of Theoretical Physics, Bijeni\v{c}ka 54 10000 Zagreb, Croatia}
\email{youjiangyang@gmail.com}

\newcommand{\tr}{\hbox{tr}}
\def\BOX{\mathord{\vbox{\hrule\hbox{\vrule\hskip 3pt\vbox{\vskip
3pt\vskip 3pt}\hskip 3pt\vrule}\hrule}\hskip 1pt}}

\date{\today}

\begin{abstract}
In this article we expound a discovery of  the quantum equivalence/duality  of U(N) noncommutative quantum field theories (NC QFT) related by the $\theta$-exact Seiberg-Witten (SW) maps and at all orders in the perturbation theory with respect to the coupling constant.  We show that this proof holds for Super Yang-Mills (SYM) theories with ${\cal N}=0,1,2,4$ supersymmetry.  In short, Seiberg-Witten map does commute with the quantization of the U(N) NCQFT independently, with or without supersymmetry.
\end{abstract}


\maketitle

\section{Introduction}

In accord with the very essence of the coupling constant perturbative description of the quantum field theory, our approach to the Seiberg-Witten map \cite{Seiberg:1999vs} issue is to build the SW map by using the expansion in terms of the coupling constant \cite{Schupp:2008fs,Horvat:2010sr,Horvat:2011iv,Horvat:2011qn,Horvat:2012vn}. 
Thus, the $\theta^{\mu\nu}$ dependence of the the coupling constant perturbative definition of the theory is treated in an exact way and, then, the UV/IR mixing effect pops up \cite{Schupp:2008fs}, inducing the noncommutative quadratic IR divergence and signaling an IR instability \cite{Armoni:2001uw}.  This IR instability can be cured in unmapped theory by making the theory supersymmetric, since supersymmetry removes the corresponding quadratic noncommutative IR divergences \cite{Zanon:2000nq, Ruiz:2000hu}. However, it was shown in \cite{Martin:2008xa} that if the noncommutative fields carry a linear realization of supersymmetry their ordinary duals under the Seiberg-Witten map carry a nonlinear realization of supersymmetry. Hence, it is far from trivial that the supersymmetry cancelation mechanism between the one-loop noncommutative quadratic IR divergences coming from bosonic and fermionic degrees of freedom works when the classical noncommutative theory is formulated, first, in terms of the ordinary fields and then quantized. And yet, it has been shown in \cite{Martin:2016zon,Martin:2016hji} that SUSY cancelation mechanism just mentioned works for all the two-point functions when we have ${\cal N}=1, 2$ and $4$ supersymmetry.

The occurrence of the UV/IR mixing phenomenon in both these quantum field theories (without and with SW map) give strong support to the idea that they are dual descriptions of the same underlying quantum field theory, at least in the perturbative  regime defined by the coupling constant. However, in the U(1) YM theory, the UV divergent part of the two-point function of the noncommutative gauge field is local, whereas the UV divergent bit of the two-point function of the ordinary theory obtained by using the $\theta$-exact SW map contains unusual $\theta$-dependent nonlocal contributions, at least in the Feynman gauge~\cite{Horvat:2011bs, Horvat:2013rga,Trampetic:2015zma,Horvat:2015aca}.  Their existence cast doubts on the truth of the quantum duality conjecture at hand. Of course, UV divergent contributions to the two-point functions are  gauge dependent.

We present here our answer(s), given in \cite{Martin:2016zon,Martin:2016hji,Martin:2016saw},
to the following 16 years old but important question:\\
-- Does the Seiberg-Witten map between different NCYMs persist after quantization?\\
Our answer will also solve the following closed related problems
\\-- Is the UV/IR mixing effect--signaling a vacuum instability--a gauge-fixing independent characteristic of both versions of theories?\\
-- Whether or not the UV divergent nonlocal terms in ordinary theory are really physically relevant?\\
-- Does SUSY help to remove the UV and IR divergences in the above NCQFT in general?

\section{Motivations}

\begin{figure}
\begin{center}
\includegraphics[width=8.5cm,angle=0]{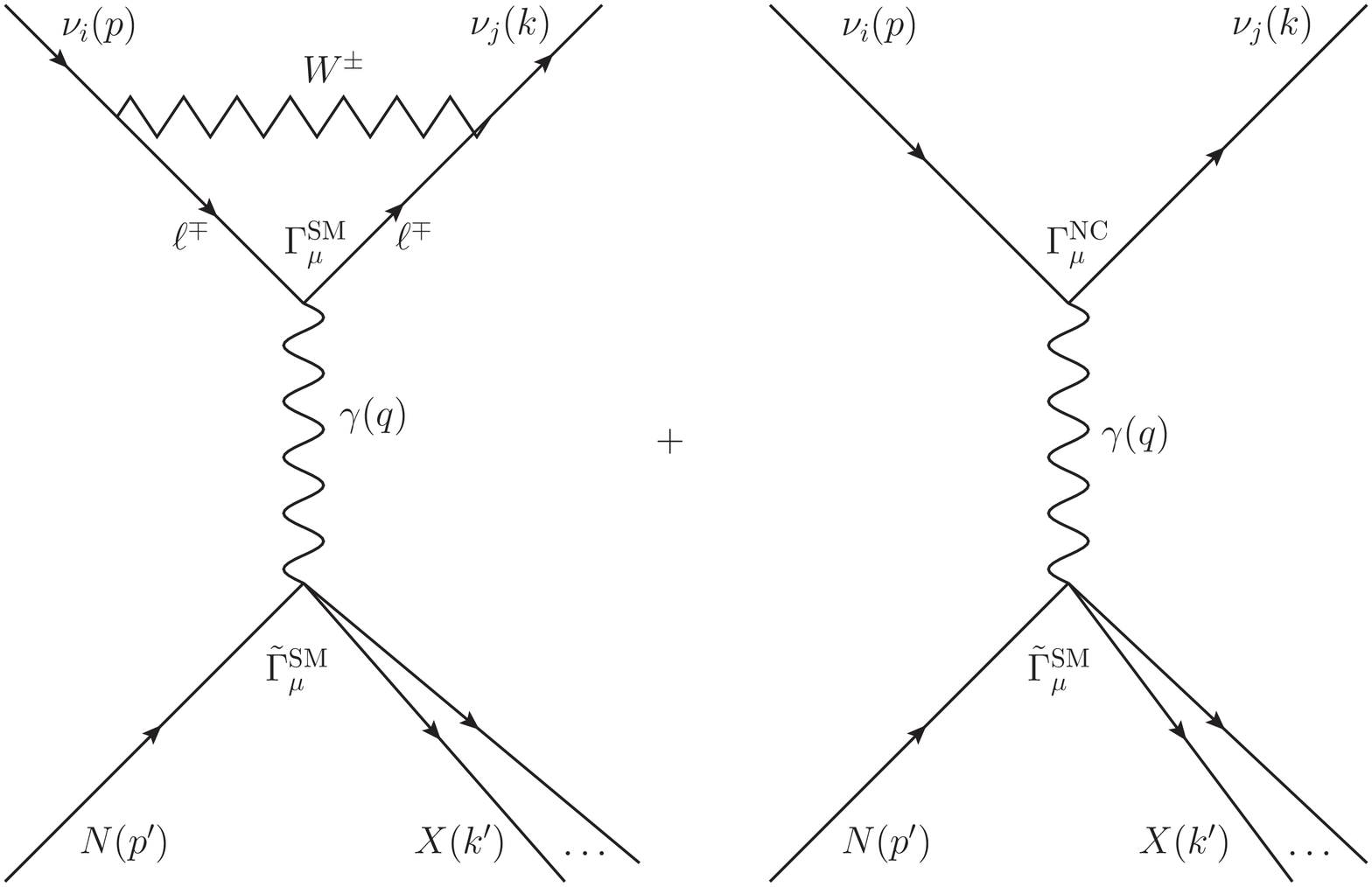}
\end{center}
\caption{Feynman diagrams giving the SM plus NC contribution to the amplitude for the $\nu +N\to\nu+ X$ process. Here
$\tilde\Gamma^{\rm SM}_\mu$ denotes textbook coupling of photons to nucleons given in \cite{Horvat:2010sr}.}
\label{Fig1}
\end{figure}

Among other, like plasmon and $Z$ forbidden and invisible decays \cite{Horvat:2011qn,Horvat:2011iv,Horvat:2012vn}, or holography  \cite{Horvat:2010km} and reheating phase after inflation \cite{Horvat:2011wh} in the framework of the noncommutativity of spacetime, our first main phenomenological motivation for using/applying the $\theta$-exact SW-map roots in physics of "the ultra-high energy neutrino  experiments", where relevant scatterings of extreme energetic neutrinos (originating from cosmic rays, so called cosmogenic neutrinos) on nucleons, as presented in our Fig. \ref{Fig1}, were analyzed in both NCQFT models, in $\theta$-expanded and $\theta$-exact models, respectively. By using the following Feynman rules rising from the relevant $\theta$-exact model \cite{Horvat:2010sr},
\begin{equation}
\begin{split}
\Gamma_{\mu}^{\rm NC}(q,k;\theta){\big|}_{\theta-{\rm exact}}&=ie(1\pm\gamma_5)\frac{\sin\frac{q\theta k}{2}}{\frac{q\theta k}{2}} V_\mu(q,k;\theta),
\\
 V_\mu(q,k;\theta){\big|}^{\rm on-shell}&=\theta_{\mu\nu\rho}q^{\nu}k^{\rho}{\big|}^{\rm on-shell}
 =\gamma_\mu (q\theta k),
 \label{FR}
\end{split}
\end{equation}
\begin{figure}
\begin{center}
\includegraphics[width=8.5cm,angle=0]{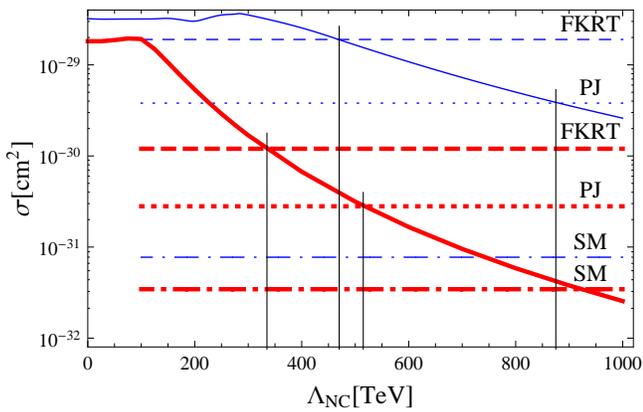}
\end{center}
\caption{The $\sigma(\nu +N\to\nu+ X)$ total cross sections
as a function of the scale of noncommutativity $\Lambda_{\rm NC}$ for $E_\nu=10^{10}$ GeV (thick lines)
and $E_\nu=10^{11}$ GeV (thin lines). FKRT and PJ lines are the upper
bounds on the neutrino--nucleon inelastic cross section, denoting different
estimates for the cosmogenic neutrino flux. SM denotes the SM total
(charged current plus neutral current) neutrino-nucleon inelastic cross
section. Vertical lines denote intersections of our curves with the
RICE results. }
\label{Fig2}
\end{figure}
we have computed Feynman diagrams from Fig. \ref{Fig1} and obtain the total cross section $\sigma(\nu+N\to\nu+X)$,
where $X$ denote anything.

Employing the upper bound on the $\nu N$ cross section derived from the RICE
Collaboration search results \cite{Kravchenko:2002mm} at $E_{\nu} =
10^{11}$ GeV ($4\times 10^{-3}$ mb for the FKRT neutrino flux \cite{Fodor:2003ph})),
in $\theta$-expanded model one can infer
that the scale of noncommutativity $\Lambda_{\rm NC}$
to be greater than 455 TeV, a really strong bound \cite{Horvat:2010sr}. One should however be
careful and suspect this result as it has been obtained from the conjecture
that the $\theta$-expansion stays well-defined in the kinematical region of
interest, and the more reliable limits on $\Lambda_{\rm NC}$ are expected to be
placed  precisely by examining low-energy processes \cite{Haghighat:2009pv}.
Although a heuristic criterion for the validity of the perturbative $\theta$-expansion,
$\sqrt{s}/\Lambda_{\rm NC} \lesssim\;1$, with $s = 2 E_{\nu}M_N$, would
underpin our result on $\Lambda_{\rm NC}$, a more thorough inspection on
the kinematics of the process does reveal a  more stronger energy
dependence  $E_{\nu}^{1/2} s^{1/4}/ \Lambda_{\rm NC} \lesssim 1$.
In spite of an additional phase-space suppression for small $x$'s in
the $\theta^2$-contribution \cite{Alboteanu:2007bp} of
the cross section relative to the $\theta$-contribution, we find
an unacceptably  large ratio $\sigma({\theta^2})/\sigma({\theta}) \simeq 10^4$,
at $\Lambda_{\rm NC}=455$ TeV. Hence,  the bound on $\Lambda_{\rm NC}$
obtained this way is incorrect, and our last resort
was to modify the model adequately to include the full-$\theta$
resummation leading to the $\theta$-exact Feynman rules (\ref{FR}), thereby allowing us to compute nonperturbatively in $\theta$. Result, presented in Fig. \ref{Fig2}, was unexpectedly successful  showing nice convergent behavior of the total cross section as a function of the scale of noncommutativity $\Lambda_{\rm NC}$ at two different extreme neutrino energies.

\begin{figure}
\begin{center}
\includegraphics[width=7cm,angle=0]{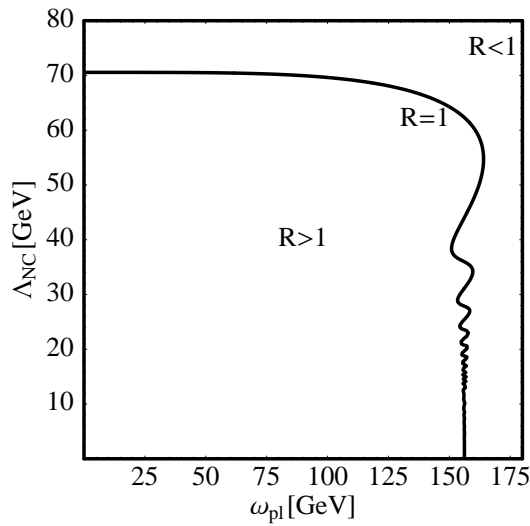}
\end{center}
\caption{The plot of the scale of noncommutativity $\Lambda_{\rm NC}$ versus
the  plasmon frequency $\omega_{\rm pl}$, obtained from plasmon decay rate $\Gamma(\gamma_{\rm pl}\to\bar\nu\nu)$. See detailes in \cite{Horvat:2011iv}.}
\label{Fig3}
\end{figure}

\begin{figure}
\begin{center}
\includegraphics[width=7cm,angle=0]{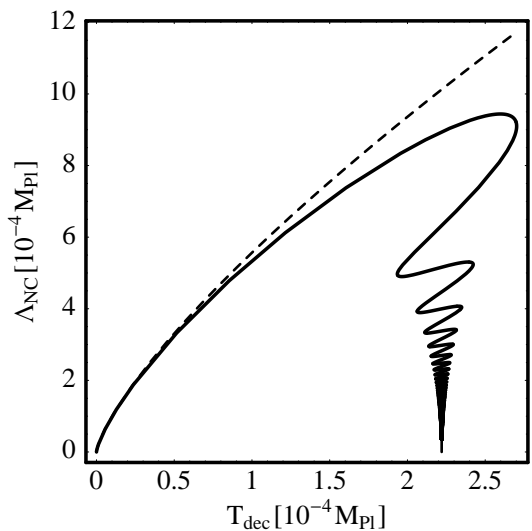}
\end{center}
\caption{The plot of the scale of noncommutativity $\Lambda_{\rm NC}$ versus
decoupling temperature $T_{\rm dec}$. The dashed/full curve corresponds to the
$\theta$-expand/$\theta$-exact solutions,
respectively. In both curves we
set, for illustration purposes, $g_{*} = g_{*}^{ch} = 100$, and the logarithmic scaling of the fine
structure constant with temperature is ignored. The full curve reveals
$T_{\rm dec}^{\rm max} = 2.7 \times 10^{-4} {\rm M_{Pl}}$ and $\Lambda_{\rm NC}^{\rm max} =
9.4 \times 10^{-4} {\rm M_{Pl}}$ \cite{Horvat:2011iv}.}
\label{Fig4}
\end{figure}

Finally we give in Fig. \ref{Fig3}, as an example of the convergent form of the physically relevant result, a plot of the scale of noncommutativity $\Lambda_{\rm NC}$ versus the  plasmon frequency $\omega_{\rm pl}$ obtained from the computation of the plasmon decay into neutrino pair rate in the neutrino mass extended NCSM  \cite{Horvat:2011iv}. In addition in Fig. \ref{Fig4}, there is a convergent plot of the scale $\Lambda_{\rm NC}$ versus BBN decoupling temperature $T_{\rm dec}$, obtained from the assumption that the plasmon decay into sterile neutrino pairs rate
$\Gamma(\gamma_{\rm pl}\to\bar\nu_R\nu_R)$ is mostly due to the noncommutativity of spacetime, see detailes in \cite{Horvat:2011iv}.

\section {Noncommutative gauge theory}
\subsection{Definitions/properties of the Moyal $\star$-product}

The Moyal space means $[\hat x^i,\hat x^j]=\theta^{ij}=\mbox{constant}$, with
$\hat x^i$ being operators.
This can be realized by the Moyal-Weyl star$-(\star)$ product
\begin{equation}
\begin{split}
f(x)\star g(x)=&\frac{1}{(2\pi)^{2n}}\int d^nkd^np\tilde f(k)\tilde g(p)
\\
&\cdot \exp[i(k_i+p_i) x^i\underbrace{-i\theta^{ij}k_ip_j}_{\rm nonlocal}]
\\
\longrightarrow [x^i\stackrel{\star}{,}x^j]=&x^i\star x^j-x^j\star x^i=i\theta^{ij},
\end{split}
\end{equation}
which is associative
and invariant under the cyclicity of  integration.

\subsection {The noncommutative $\rm U(1)$ gauge action}

The pure gauge action $S_g$
\begin{gather*}
S_g=-\frac{1}{4}\int F_{\mu\nu}\star F^{\mu\nu},
\\
F_{\mu\nu}=\partial_\mu A_\nu-\partial_\nu
A_\mu-i[A_\mu\stackrel{\star}{,}A_\nu],
\\
\delta_\Lambda
A_\mu=\partial_\mu\Lambda+i[\Lambda\stackrel{\star}{,}A_\mu],\;
\delta_\Lambda F_{\mu\nu}=i[\Lambda\stackrel{\star}{,}F_{\mu\nu}],
\end{gather*}
\normalsize
is interacting, yet perturbatively workable:
\begin{equation*}
\begin{split}
S_g&=-\frac{1}{4}\int \underbrace{(\partial_\mu A_\nu-\partial_\nu
A_\mu)^2}_{\rm undeformed\; free\; part}
\\&\underbrace{-2i(\partial_\mu A_\nu-\partial_\nu
A_\mu)[A^\mu\stackrel{\star}{,}A^\nu]-[A_\mu\stackrel{\star}{,}A_\nu]^2}_{\rm non-local\;interaction}.
\end{split}
\end{equation*}
Introducing Moyal product turns the commutative $\rm U(1)$ theory, which is a free theory, into the NC $\rm U(1)$ gauge theory which is a nonlocally interacting theory. Also, due to the cyclicity, field theories on Moyal space admits relatively simple pertubative quantization.\\
Nonlocal factor regularize part of the Schwinger parameterized loop integral, turning it into integral over modified bessel function $K_n$'s which is IR divergent.
$K_n$'s contain logarithmic function in its expansion, which could cause unitarity problem when performing Wick rotation. This study is restricted to the Euclidean space only.

\subsection{Supersymmetric U(1) NCQFT}
Classical SUSY is easy to realize on Moyal space:
\begin{equation}
\begin{split}
S_{{\cal N} =1}&= S_g
+ i \bar{\Lambda}_{\dot{\alpha}}\bar{\sigma}^{\mu\,
\dot{\alpha}\alpha}{\cal D}_\mu[A]\Lambda_{\alpha}+\frac{1}{2} D^{(nc)}D^{(nc)},
\end{split}
\label{2.1}
\end{equation}
since the SUSY transformations are linear:
\begin{equation}
\begin{split}
&\delta_{\xi}\Lambda_{\alpha}=-iD^{(nc)}\xi_{\alpha}-e^{-1}(\sigma^{\mu}\bar{\sigma}^{\nu})_{\alpha}^{\phantom{\alpha}\beta}\xi_\beta F_{\mu\nu},
\\&\delta_{\xi}A^{\mu}=ie(\xi\sigma^{\mu}\bar{\Lambda}-\Lambda\sigma^{\mu}\bar{\xi}),
\\&\delta_{\xi}D^{(nc)}=(\xi\sigma^{\mu}{\cal D}_{\mu}\bar{\Lambda}-{\cal D}_{\mu}\Lambda\sigma^{\mu}\bar{\xi}),
\end{split}
\label{2.2}
\end{equation}
and there exists superfield formalism.

Extended SUSY actions can also be constructed, as given below: 
\begin{equation}
\begin{split}
&S_{{\cal N}=2}=S_g
+({\cal D}_\mu[A]\Phi)^{\dagger} {\cal D}^\mu[A] \Phi
\\&-
\frac{e^2}{2}[\Phi^{\dagger}\stackrel{\star}{,}\Phi]^2
+i  \bar{\Lambda}\bar{\sigma}^{\mu}{\cal D}_\mu[A]\Lambda
\\&
+i  \bar{\Psi}\bar{\sigma}^{\mu}{\cal D}_\mu[A]\Psi
+ie\sqrt{2}\Psi[\Lambda\stackrel{\star}{,}\Phi^{\dagger}]
+ ie\sqrt{2}\bar{\Psi}[\bar{\Lambda}\stackrel{\star}{,}\Phi],
\end{split}
\label{sn2action}
\end{equation}

\begin{equation}
\begin{split}
S_{{\cal N}=4}&=S_g
+i  \bar{\Lambda}^{i}\bar{\sigma}^{\mu}{\cal D}_\mu[A]\Lambda_{i}
\\&
+
\frac{1}{2}{\cal D}_\mu[A] \Phi_m{\cal D}^\mu[A]\Phi_m
+\Big(\frac{e}{2}[\Phi_m\stackrel{\star}{,}\Phi_n]\Big)^2
\\&
+i\frac{e}{2}(\tilde{\sigma}^{-1})^{ij}\Lambda_i[\Lambda_j\stackrel{\star}{,}\Phi_m]
-i\frac{e}{2}(\tilde{\sigma})_{ij}\bar{\Lambda}^i[\bar{\Lambda}^j\stackrel{\star}{,}\Phi_m].
\end{split}
\label{sn4action}
\end{equation}

\subsection{Photon polarization tensor computed in the Feynman gauge}

The one-loop quantum corrections in the NC $\rm U(1)$ gauge theory on Moyal space has the structure where the UV, quadratic and logarithmic IR divergences coexist. Photon polarization tensor computed in the pure $\rm U(1)$ with Feynman gauge shows UV and IR divergences: 
\begin{equation}
\begin{split}
\Pi_{\rm photon}^{\mu\nu}&(p)\sim \frac{g^2}{(4\pi)^2}\Bigg(\frac{10}{3}(g^{\mu\nu}p^2-p^\mu p^\nu)
\\&\cdot\Big(\frac{2}{\epsilon}
\underbrace{-\ln \frac{p^2}{\mu^2}+\ln (p^2(\theta p)^2)}_{+\ln (\mu^2(\theta p)^2)}\Big)
+32\underbrace{\frac{(\theta p)^{\mu}(\theta p)^{\nu}}{(\theta p)^4}}_{\rm quadratic}\bigg),
\end{split}
\end{equation}
The $\beta$-function obtained from the planar UV divergences in the photon two and three point functions suggests asymptotic freedom.  Yet the quadratic IR divergence is an even bigger concern,
while logarithmic term is actually the one that really mixes with the UV term.

Question is wether supersymmetry could help? Answer is yes since SUSY controls IR divergences  \cite{Martin:2016zon}.
For the $\mathcal N=1$ SUSY suppresses the quadratic IR divergence,
\begin{equation}
\begin{split}
\Pi_{\rm \mathcal N=1}^{\mu\nu}&(p)=\Pi_{\rm photon}^{\mu\nu}(p)+\Pi_{\rm photino}^{\mu\nu}(p)
\\ \sim \frac{g^2}{(4\pi)^2}&\Bigg(\bigg(\frac{10}{3}\big(g^{\mu\nu}p^2-p^\mu p^\nu\big)
\Big(\frac{2}{\epsilon}+\ln(\mu^2(\theta p)^2)\Big)
\\&+32\frac{(\theta p)^{\mu}(\theta p)^{\nu}}{(\theta p)^4}\bigg)-\bigg(\frac{4}{3}\big(g^{\mu\nu}p^2-p^\mu p^\nu\big)
\\&\cdot\Big(\frac{2}{\epsilon}+\ln(\mu^2(\theta p)^2)\Big)
+32\frac{(\theta p)^{\mu}(\theta p)^{\nu}}{(\theta p)^4}\bigg)\Bigg),
\end{split}
\end{equation}
while the
$\mathcal N=4$ SUSY renders the theory finite. This we shall give explicitly at the end of this article.

\section{Quantum duality of QFT's related by the $\theta$-exact SW map}

Classical noncommutative field theories admit an equivalent representation in terms of ordinary fields formulated by employing the Seiberg-Witten (SW) map~\cite{Seiberg:1999vs}. However we still do not know whether this equivalence holds at the quantum level, i.e., whether the quantum theory defined in terms of the NC fields is the same as the quantum theory defined in terms of commutative fields, and obtained from the NC action by using the $\theta$-exact SW map~\cite{Seiberg:1999vs}. In this article we prove that the $\theta$-exact  Seiberg-Witten map establishes an equivalence relation between perturbative --in the coupling constant-- quantum field theories defined with respect to the noncommutative and commutative fields, by showing that the corresponding on-shell DeWitt effective actions~\cite{DeWitt:1967ub,Kallosh:1974yh,DeWitt:1980jv,DeWitt:1988fm} can be SW-mapped one to each other \cite{Martin:2016hji,Martin:2016saw}. We also give an explicit check of our verdict in the (supersymmetric) NC $\rm U(1)$ gauge theory.

This result gives further robustness to the quantum duality conjecture between the formulation in terms of ordinary fields and the description in terms of noncommutative fields. However, the nonlocal UV divergent structure still persists after introducing supersymmetry into the game. But, by using two different gauge-fixing terms, it was shown in  \cite{Martin:2016zon} that the nonlocal  UV divergent contributions are gauge dependent and, therefore, it could be possible to remove them. This is unlike the noncommutative quadratic IR divergences which do not change with the gauge-fixing term as was proved in \cite{Ruiz:2000hu}, in the noncommutative field description, and in \cite{Martin:2016zon}, in the ordinary field formulation, respectively.

\subsection{DeWitt effective action in the path integral formulation}
The on-shell DeWitt effective action \cite{DeWitt:1988fm} with respect to the noncommutative/hatted fields, $\hat\Gamma_{\rm DeW}\big[\hat B_\mu\big]$, is given by the following path integral formulation
\begin{equation}
\begin{split}
&e^{\frac{i}{\hbar}\hat\Gamma_{\rm DeW}\big[\hat B_\mu\big]}=\int d\hat Q_\mu^a d\hat C^a d\hat{\bar C}^a d\hat F^a\;
\\&\cdot e^{\frac{i}{\hbar}S_{\rm NCYM}\big[\hat B_\mu+\hbar^{\frac{1}{2}}\hat Q_\mu\big]+i S_{\rm gf}\big[\hat B_\mu,\hat Q_\mu,\hat F,\hat{\bar C},\hat C\big]},
\end{split}
\label{effectivenoncommutative}
\end{equation}
where $S_{\rm NCYM}=-\frac{1}{4g^2}\int \tr\,\hat F_{\mu\nu}\hat F^{\mu\nu}$ is the usual NC $\rm U(N)$ Yang-Mills (YM) action, while $S_{\rm gf}$ is the gauge-fixing action which can be expressed in the BRST language as
\begin{equation}
S_{\rm gf}\big[\hat B_\mu,\hat Q_\mu,\hat F,\hat{\Bar C},\hat C\big]=\hat\delta_{BRS}\,X_{\rm gf}\big[\hat B_\mu,\hat Q_\mu,\hat F,\hat{\Bar C},\hat C\big].
\label{Xgf}
\end{equation}
Above $X_{\rm gf}\big[\hat B_\mu,\hat Q_\mu,\hat F,\hat{\Bar C},\hat C\big]$ is an arbitrary gauge-fixing functional. The noncommutative
$\rm U(N)$ BRS transformations $\hat\delta_{BRS}\hat A_\mu=\hat D_\mu\hat C$ and $\hat\delta_{BRS}\hat C=-i\hat C\star\hat C$ induce the following BRS transformations after introducing the noncommutative background-field splitting
$\hat A_\mu\to \hat B_\mu+\hbar^{\frac{1}{2}}\hat Q_\mu$,

\begin{equation}
\begin{array}{cr}
\hat\delta_{\rm BRS}\hat B_\mu=0,\;\hat\delta_{\rm BRS}\hat Q_\mu=\hbar^{-\frac{1}{2}} \hat D_\mu\big[\hat B_\mu+\hbar^{\frac{1}{2}}\hat Q_\mu\big]\hat C,
\\
\hat\delta_{\rm BRS}\hat C=-i\hat C\star\hat C,\;
\hat\delta_{\rm BRS}\hat{\bar C}=\hbar^{-\frac{1}{2}}\hat F,\;
\hat\delta_{\rm BRS}\hat F=0.
\end{array}
\label{hatBRS}
\end{equation}

The $\theta$-exact SW map of the NC fields in terms of commutative/ordinary fields in the $\rm U(N)$ gauge theory
\begin{gather}
\hat A_\mu=\hat A_\mu\left[A_\mu,\theta\right],\:\;
\hat C=\hat C\left[A_\mu,C,\theta\right],
\label{SW}
\end{gather}
are solutions to the following equations
\begin{gather}
\hat\delta_{\rm BRS}\hat A_\mu=\delta_{\rm BRS}\hat A_\mu\left[A_\mu,\theta\right],\:\:
\hat\delta_{\rm BRS}\hat C=\delta_{\rm BRS}\hat C\left[A_\mu,C,\theta\right].
\label{SWeqs}
\end{gather}
They can be expressed $\theta$-exactly as formal power series of the field operators~\cite{Martin:2012aw, Martin:2015nna}:
\begin{gather}
\hat A_\mu\left[A_\mu,\theta\right](x)=A_\mu(x)+\sum\limits_{n=2}^\infty \mathcal{A}_\mu^{(n)}(x),
\label{SW1}
\\
\hat C\left[A_\mu,C,\theta\right](x)=C(x)+\sum\limits_{n=1}^\infty \mathcal{C}^{(n)}(x),
\label{SW2}
\end{gather}
where
\begin{gather}
\begin{split}
\mathcal{A}_\mu^{(n)}(x)=\int&\prod\limits_{i=1}^n\frac{d^4 p_i}{(2\pi)^4} e^{i\left(\sum\limits_{i=1}^n p_i\right)x}
\\&\cdot\mathfrak{A}^{(n)}_\mu\big[(a_1,\mu_1,p_1),......,(a_n,\mu_n,p_n);\theta\big]
\\&\cdot\tilde A_{\mu_1}^{a_1}(p_1)......\tilde A_{\mu_n}^{a_n}(p_n),
\end{split}
\label{SW3}
\\
\begin{split}
\mathcal{C}^{(n)}(x)&=\int\prod\limits_{i=1}^n\frac{d^4 p_i}{(2\pi)^4} e^{i\left(p+\sum\limits_{i=1}^n p_i\right)x}
\\&\cdot\mathfrak{C}^{(n)}\big[(a_1,\mu_1,p_1),......,(a_n,\mu_n,p_n);(a,p);\theta\big]
\\&\cdot \tilde A_{\mu_1}^{a_1}(p_1)......\tilde A_{\mu_n}^{a_n}(p_n)C^a(p).
\end{split}
\label{SW4}
\end{gather}
The quantities $\mathfrak{A}_\mu^{(n)}$ and $\mathfrak{C}^{(n)}$ are totally symmetric under the permutations with respect to the set of the parameter-triples $\left\{(a_i,\mu_i,p_i)|i=1,...,n\right\}$, which have the property --of key importance-- that only the momenta which are not contracted with $\theta^{\mu\nu}$ build up polynomials which  never occur in the denominator \cite{Martin:2012aw, Martin:2015nna}.

Using the ordinary background-field splitting
\begin{equation}
A_\mu=B_\mu+\hbar^{\frac{1}{2}}Q_\mu,
\label{ordinarysplit}
\end{equation}
with the corresponding BRS transformations,
\begin{equation}
\delta_{\rm BRS} B_\mu=0,\; \delta_{\rm BRS} Q_\mu=\hbar^{-\frac{1}{2}} D_\mu\big[B_\mu+\hbar^\frac{1}{2} Q_\mu\big]C,
\label{normalBRS}
\end{equation}
where $B_\mu$ is the commutative background field and $Q_\mu$ the commutative quantum fluctuation,  for the SW map \eqref{SW} we find the background-field splitting
\begin{equation}
\begin{split}
\hat A_\mu\big[B_\mu+\hbar^{\frac{1}{2}}Q_\mu,\theta\big]=&\hat A_\mu\big[B_\mu,\theta\big]+\hbar^{\frac{1}{2}}\hat Q_\mu\big[B_\mu,Q_\mu,\hbar,\theta\big]
\\=&\hat B_\mu\big[B_\mu,\theta\big]+\hbar^{\frac{1}{2}}\hat Q_\mu\big[B_\mu,Q_\mu,\hbar,\theta\big]
\\
\hat C\big[B_\mu+\hbar^{\frac{1}{2}}Q_\mu,C,\theta\big]&=\hat C\big[B_\mu,C,\theta\big]
\\&+\hbar^{\frac{1}{2}}\hat C^{(1)}\big[B_\mu,Q_\mu,C,\hbar,\theta\big],
\end{split}
\label{Qdefinition}
\end{equation}
which ensures that the ordinary BRS transformations \eqref{normalBRS} induces the NC BRS transformations \eqref{hatBRS}.

Now the on-shell DeWitt action with respect to the ordinary fields, $\Gamma_{\rm DeW}\left[B_\mu\right]$ is given by the path integral
\begin{equation}
\begin{split}
&e^{\frac{i}{\hbar}\Gamma_{\rm DeW}\big[B_\mu\big]}=\int d Q_\mu^a d C^a d \hat{\bar {C^a}} d \hat F^a\;
\\&
\cdot e^{\frac{i}{\hbar}S_{\rm NCYM}\big[B_\mu+\hbar^{\frac{1}{2}}Q_\mu\big]+i  S_{\rm gf}\big[ B_\mu, Q_\mu,\hat F,\hat{\bar C},C\big]},
\end{split}
\label{effectiveaction}
\end{equation}
in which we change variables: $C^a \to \hat C^a$ and $Q^a\to\hat Q^a$, so that it transforms into the new path integral
\begin{eqnarray}
&&{e^{\frac{i}{\hbar}\Gamma_{\rm DeW}\big[B_\mu\big]}=\int d\hat Q_\mu^a d \hat C^a d\hat{\bar C}^a d \hat F^a\;J^{-1}_1[B,Q]\,J_2[B,Q]\,}\nonumber\\
&&{\cdot e^{\frac{i}{\hbar}S_{\rm NCYM}\big[\hat B_\mu+\hbar^{\frac{1}{2}}\hat Q_\mu\big]+i  S_{\rm gf}\big[\hat B_\mu, \hat Q_\mu,\hat F,\hat{\bar C},\hat C\big]},}
\label{effectiveactionchanged}
\end{eqnarray}
containing the Jacobian determinants $J_1\big[B^a,Q^a\big]$ and $J_2\big[B^a,Q^a\big]$ who are defined as follows
\begin{equation}
\begin{array}{l}
{J_1\big[B^a, Q^a\big]\,=\,\det\frac{\delta\hat Q^a_\mu(x)}{\delta Q^b_\nu(y)}\,=\,\exp\,{\rm Tr}\,\ln\Big(\frac{\delta\hat Q^a_\mu(x)}{\delta Q^b_\nu(y)}\Big)},\\[8pt]
{ J_2\big[B^a,Q^a\big]\,=\,\det\frac{\delta\hat C^a(x)}{\delta C^b(y)}\,=\,\exp\,{\rm Tr}\,\ln\Big(\frac{\delta\hat C^a(x)}{\delta C^b(y)}\Big).}
 \end{array}
 \label{jacobdet}
 \end{equation}

 \subsection{Triviality of the Jacobian determinants}

Under the assumption that both above Jacobians are equal to one,
we can prove that the right hand side of (\ref{effectiveactionchanged}) equals to the right hand side of (\ref{effectivenoncommutative}), so that
\begin{equation}
\Gamma_{\rm DeW}\big[B_\mu\big]=\hat\Gamma_{\rm DeW}\big[\hat B_\mu[B_\mu]\big].
\label{equiv}
\end{equation}

Note that above result is valid on-shell, i.e. when $\hat B_\mu[B_\mu]$ satisfies the NC YM equations of motions
\begin{equation}
\hat D_{\mu}\big[\hat B_\mu[B_\mu]\big]\hat F^{\mu\nu}\big[\hat B_\mu[B_\mu]\big]=0,
\label{NCYMEOM}
\end{equation}
and the reason is the on-shell uniqueness of DeWitt effective action~\cite{Kallosh:1974yh,Ichinose:1992np}.

Using the SW map expansion \eqref{SW3} and the background-field splitting (\ref{Qdefinition}) one can show that
\begin{equation}
\begin{array}{l}
{\frac{\delta\hat Q^a_\mu(x)}{\delta Q^b_\nu(y)}\,=\,\frac{1}{\hbar^{\frac{1}{2}}}\,\frac{\delta\hat A^a_\mu(x)}{\delta Q^b_\nu(y)}
=\delta^a_b\delta^\nu_\mu\,\delta(x-y)
+\sum\limits_{n=2}^{\infty}\,
\int\prod\limits_{i=1}^n\frac{d^4 p_i}{(2\pi)^4}}
\\[8pt]{\cdot
e^{i\left(\sum\limits_{i=1}^{n-1} p_i\right)x}\,e^{ip_n (x-y)}
{\cal M}^{(n)\,a\,\nu}_{\phantom{(n)\,}b\,\mu}(p_1,p_2,....p_{n-1};p_n;\theta)},
\end{array}
\label{dethQQ}
\end{equation}
where
\begin{equation}
\begin{split}
&{\cal M}^{(n)\,a\,\nu}_{\phantom{(n)\,}b\,\mu}(p_1,p_2,....p_{n-1};p_n;\theta)
\\
&=n\;\tr\Big[T^a{\mathfrak{A}^{(n)}_\mu\big[(a_1,\mu_1,p_1)},
...,(a_{n-1},\mu_{n-1},p_{n-1}),
\\&
(b,\nu,p_n);\theta\big]\Big]
\tilde A_{\mu_1}^{a_1}(p_1)...\tilde A_{\mu_{n-1}}^{a_{n-1}}(p_{n-1}).
\end{split}
\label{Mcaldef}
\end{equation}
Note  that $\tilde A_{\mu_i}^{a_i}(p_i)=\tilde B_{\mu_i}^{a_i}(p_i)+\hbar^{\frac{1}{2}}\tilde Q_{\mu_i}^{a_i}(p_i)$ for all $i$.

Let $l_i$, $i=1,..,m+1$ be given by
\begin{equation*}
l_1=\sum\limits_{i_1=1}^{n_1-1}\, p_{1,i_1},
\:...........\:, l_{m+1}=\sum\limits_{i_{m+1}=1}^{n_{m+1}}\,p_{m+1,i_{m+1}},
\end{equation*}
then, by taking into account (\ref{dethQQ}) and carrying out a lengthy straightforward computation one gets
\begin{equation}
\begin{array}{l}
{\ln\,J_1[B,Q]={\rm Tr}\ln\,\Big(\frac{\delta\hat Q^a_\mu(x)}{\delta Q^b_\nu(y)}\Big)={\sum\limits_{n=2}^{\infty}}
{\int\prod\limits_{i=1}^{n-1}\frac{d^4 p_i}{(2\pi)^4}\,\delta\Big(\sum\limits_{i=1}^{n-1}p_i\Big)}}
\\[8pt]
{\cdot\int\frac{d^4 q}{(2\pi)^4}\,
{\cal M}^{(n)\,a\,\mu}_{\phantom{(n)\,}a\,\mu}\left(p_1,p_2,....,p_{n-1};q;\theta\right)+\sum\limits_{m=1}^{\infty}\frac{(-1)^{m}}{m+1}}
\\[8pt]
{\cdot\sum\limits_{n_1=2}^{\infty}\cdots\sum\limits_{n_{m+1}
=2}^{\infty}
\int\prod\limits_{i_1=1}^{n_1-1}\frac{d^4 p_{1,i_1}}{(2\pi)^4}\cdots\int\prod\limits_{i_{m+1}=1}^{n_{m+1}-1}\frac{d^4 p_{m+1,i_{m+1}}}{(2\pi)^4}}
\\[8pt]
{\cdot\delta\Big(\sum\limits_{i=1}^{m+1}l_i\Big)\int\frac{d^4 q}{(2\pi)^4}\Big[
{\cal M}^{(n_1)\,a\,\mu_1}_{\phantom{(n_1)\,}a_1\,\mu}\left(p_{1,1},p_{1,2},....,p_{1,n_1-1};q;\theta\right)
}\\[8pt]
{\cdot{\cal M}^{(n_2)\,a_1\,\mu_2}_{\phantom{(n_3)\,}a_2\,\mu_1}\left(p_{2,1},p_{2,2},....,p_{2,n_2-1};q-l_2;\theta\right)}
\\[8pt]
{\cdots\cdots\cdots}
\\[8pt]
{\cdot{\cal M}^{(n_{m+1})\,a_m\,\mu}_{\phantom{(n_{m+1}\,}a\,\,\,\mu_m}\Big(p_{m+1,1},p_{m+1,2},
....,p_{m+1, n_{m+1}-1,}}
\\[8pt]
{q-\sum\limits_{i=2}^{m+1}l_i;\theta\Big)\Big].}
\label{genexpln}
\end{array}
\end{equation}
\begin{figure}
\begin{center}
\includegraphics[width=8.5cm]{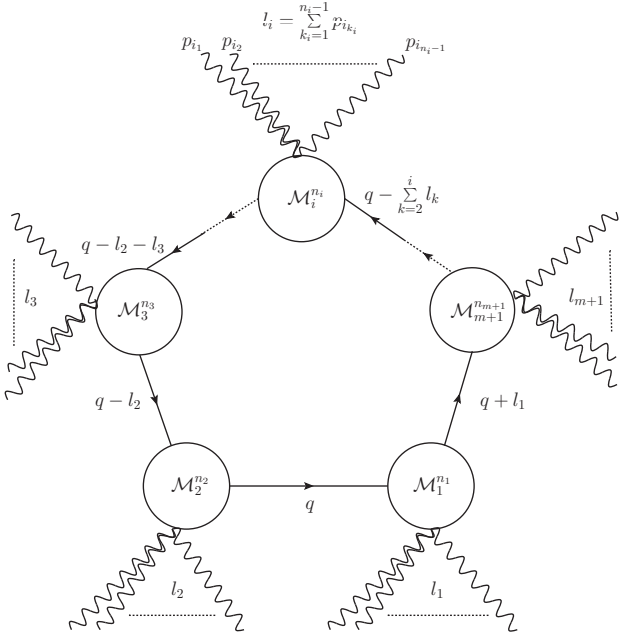}
\end{center}
\caption{The one-loop diagram interpretation/ilustration of (\ref{genexpln}): Each circle corresponds one $\mathcal M^{(n_i)}$, wavy lines denote the gauge field operators, either background or quantum, within the $\mathcal M^{(n_i)}$. The $l_i$'s are then just the total momentum brought in by these field operators. The solid line flows in each circle gives the assignment of $q-\sum_k l_k$ into the corresponding $\mathcal M^{(n_i)}$ in (\ref{genexpln}).}
\label{fig:InterpretTrlnJ1}
\end{figure}
The general structure of the master integral (\ref{genexpln}) above can be visualized as a kind of one-loop diagram, given in Fig.~\ref{fig:InterpretTrlnJ1}.

Hence, in view of the above equations (\ref{Mcaldef}) and (\ref{genexpln}), to compute $\ln\,J_1[B,Q]$ one has to work out the following dimensionally regularized type of integrals over the internal momenta $q^\mu$:
\begin{equation}
\begin{array}{l}
{\mathfrak{V}=\int\frac{d^D q}{(2\pi)^D}\Big\{
{\tr\Big[T^a{\mathfrak{A}_\mu^{(n_1)}}\big[(b_{1,1},\nu_{1,1},p_{1,1}),}}
\\[8pt]
{.....,(b_{1,n_1-1},\nu_{1,n_1-1},p_{1,n_1-1}),(a_1,\mu_1,q);\theta\big]\Big]}
\\[8pt]
{\cdots\cdots\cdots}
\\[8pt]
{\cdot\tr\Big[T^{a_m}{\mathfrak{A}^{(n_{m+1})}}_{\mu_m}\big[(b_{m+1,1},\nu_{m+1,1},p_{m+1,1}),....,}
\\[8pt]
{(b_{m+1,n_{m+1}-1},\nu_{m+1,n_{m+1}-1},p_{m+1,n_{m+1}-1}),}
\\[8pt]
{(a,\mu,q-\sum\limits_{i=2}^{m+1} l_i);\theta\big]\Big]\Big\}.}
\label{dimregint}
\end{array}
\end{equation}
However, the previous integral in (\ref{dimregint}) is a linear combination of integrals of the type
\begin{equation}
\mathfrak{I}\,=\,\int\frac{d^D q}{(2\pi)^D}\,\mathbb{Q}(q)\,\mathbb{I}(q\theta k_i,k_i\theta k_j),
\label{mathI}
\end{equation}
where $\mathbb{Q}(q)=q^{\rho_1}q^{\rho_2}q^{\rho_3}\cdots $, $q\theta k_i=q_\mu\theta^{\mu\nu}k_{i\nu}$ and $k_i\theta k_j=k_{i\mu}\theta^{\mu\nu}k_{j\nu}$, $\forall k_{i,j}\not= q$. It is important to stress that $\mathbb{Q}(q)$ is a monomial on $q^{\rho}$ and that the functional $\mathbb{I}$, as indicated in the integrand of the integral (\ref{mathI}), is a function of the variables $q\theta k_i$ and $k_i\theta k_j$ only, and, hence, as shown in details in \cite{Martin:2016saw}, one concludes that
\begin{equation}
\mathfrak{I}=0\: \,\,\to\:\,\, \mathfrak{V}=0,
\label{22}
\end{equation}
under dimensional regularization \cite{Collins:1984xc}.
By substituting $\mathfrak{V}=0$ in (\ref{genexpln}), we obtain that in dimensional regularization the following result holds
\begin{equation}
\ln\,J_1[B,Q]=0;
\label{dimregJ1}
\end{equation}
proving that indeed $J_1[B,Q]=1$.

It is straightforward to see that identical arguments apply to $J_2[B,Q]$ as well, thus the Seiberg-Witten map equivalence between quantum theories defined in terms of noncommutative fields and in terms of ordinary fields indeed holds up to all orders in the perturbation theory.

 Now, since the $\theta$-exact Seiberg-Witten map for matter fields --see \cite{Martin:2012aw, Martin:2015nna}-- have expressions analogous to that of the ghost field, it is clear that the Jacobian of the transformation from ordinary matter fields to noncommutative matter fields is also trivial in dimensional regularization. Hence, the conclusion that we have reached above when no matter fields are included remains valid when the latter are included: the on-shell De Witt action of the theory defined in terms of noncommutative fields is the same as the on-shell DeWitt action of the ordinary theory obtaiend by using the $\theta$-exact Seiberg-Witten map.

 \subsection{Two-point functions in the background field gauge}

We have checked the  equivalence established above by computing the one-loop quantum correction to the quadratic part of the effective action of the U(1) NCGFT in the NC background-field gauge prior to and after the Seiberg-Witten map. In this specific case the general equivalence reduces to a simple relation
\begin{equation}
\hat\Pi^{\mu\nu}(p)=\Pi^{\mu\nu}(p)\Big|_{\rm on-shell}.
\label{on-shell}
\end{equation}

The standard procedure for computing DeWitt effective action of the NC $\rm U(1)$ gauge theory perturbatively in the background-field formalism~\cite{Kallosh:1974yh,DeWitt:1980jv} evaluates 1-PI diagrams with all background-field external legs and all integrand field ($\hat Q_\mu,\hat{\bar C},\hat C,\hat F$) internal lines using the following action:
\begin{equation}
\begin{split}
\hat S_{\rm loop}=&S_{\rm BFG}+S_{\rm NCYM}\big[\hat B_\mu+\hat Q_\mu\big]-S_{\rm NCYM}\big[\hat B_\mu\big]
\\&-\int \bigg(\frac{\delta}{\delta \hat B_\mu}S_{\rm NCYM}\big[\hat B_\mu\big]\bigg)\cdot\hat Q_\mu.
\end{split}
\label{NCBFG}
\end{equation}
We choose $\theta$-exact SW map from $\hat S_{\rm loop}\to S_{\rm loop}$,  and then use the resulted action
\begin{equation}
\begin{split}
&S_{\rm loop}=S_{\rm BFG}\big[B_\mu,Q_\mu,\hat{\bar C}, C,\hat F\big]
\\&+S_{\rm NCYM}\Big[\hat B_\mu\big[B_\mu\big]+\hat Q_\mu\big[Q_\mu,B_\mu\big]\Big]
-S_{\rm NCYM}\Big[\hat B_\mu\big[B_\mu\big]\Big]
\\&-\int \bigg(\frac{\delta }{\delta \hat B_\mu}S_{\rm NCYM}\big[\hat B_\mu\big]\bigg)\big[B_\mu\big]\,\cdot\,\hat Q_\mu\big[B_\mu,Q_\mu\big],
\end{split}
\label{SWBFG}
\end{equation}
for one-loop computation of the effective action with respect to the ordinary fields. This choice can be shown to be equivalent to the subtraction of commutative equations of motions
$\frac{\delta}{\delta B_\mu}S_{\rm NCYM}\big[\hat B_\mu[B_\mu]\big]=0$ on-shell as long as the Seiberg-Witten map is invertible.

In the follow-on computation, by using the extended version of dimensional regularization scheme~\cite{Martin:2016zon}, we find that photon  one loop 1-PI two point functions from \eqref{NCBFG} and \eqref{SWBFG} turns out to be actually exactly the same:
\begin{equation}
\hat\Pi^{\mu\nu}(p)=\Pi^{\mu\nu}(p).
\label{hatPi}
\end{equation}
Owing to the fact that $\theta$-exact Seiberg-Witten map for the matter fields is analogous to that of the ghost field \cite{Martin:2012aw, Martin:2015nna},  the same conclusion as above (\ref{hatPi}) holds for the super partners, i.e.
\begin{equation} 
\hat\Gamma^{\dot\alpha\alpha}(p)=\Gamma^{\dot\alpha\alpha}(p),\;\;
\hat\Gamma_{(\phi)}(p)=\Gamma_{(\phi)}(p),
\label{hatGamma}
\end{equation}
which altogether verifies the equivalence relation \eqref{equiv} \footnote{Explicit computations of the photon polarization tensor $\Pi^{\mu\nu}$, as well as the super partners two point functions, with full technical details are presented in~\cite{Martin:2016zon,Martin:2016saw}.}.

 As a consequence of (\ref{hatPi}) of importance is that once we turn on supersymmetry~\cite{Martin:2016zon} both, the photon polarization tensor IR and UV cancellation results
\begin{gather}
\begin{split}
\Pi^{\mu\nu}_{\rm BFGtotal} \big|_{\rm UV}&=\frac{g^2}{(4\pi)^2}\Big(\frac{22}{3}-\frac{4}{3}{\rm N_f}-\frac{1}{3}{\rm N_s}\Big)
\\&\cdot\big(g^{\mu\nu}p^2-p^\mu p^\nu\big)\Big(\frac{2}{\epsilon}+\ln(\mu^2(\theta p)^2)\Big),
\\
\Pi_{\rm total}^{\mu\nu}\big|_{\rm IR}&=\frac{g^2}{(4\pi)^2}\Big(32-32 {\rm N_f} + 16 {\rm N_s}\Big)\frac{(\theta p)^\mu(\theta p)^\nu}{(\theta p)^4},
\end{split}
\label{N=1,2,4sum}
\end{gather}
found prior to the Seiberg-Witten map now hold precisely after the Seiberg-Witten map.

\section{Discussion}

In this presentation we state a result concerning the
equivalence of two formulations of noncommutative quantum field
theories. On the one hand there is the intrinsic formulation of
noncommutative U(N) gauge theory, and on the other hand there is a
re-formulation using only commutative fields via the SW
map, which is a change of field variables. We claim that
these two formulations, using their respective path integral
quantization, lead to perturbatively equivalent quantum field
theories, or in other words the SW map commutes with quantization.
The non-trivial part of this claim is that certain Jacobians in this
change of field variables are trivial. 
We extend the SW map valid for classical
noncommutative U(N) gauge fields $A_\mu \to \hat A_\mu$ and
corresponding ghost fields $C \to \hat C $. Then we compare on-shell
DeWitt actions $\Gamma_{\rm DeW}[B_\mu]$ and $\hat\Gamma_{\rm DeW}[\hat{B}_\mu]$ given by functional integrals. 
Both functional integrals are related by Seiberg-Witten
map. We show that both expressions are equal to all orders in
coupling constant perturbation theory. This duality holds also for
Super Yang-Mills theory with $\cal N $=4. 

We proved remarkable main result that
the perturbative gauge theory in NC space derived from classical
fields $\hat A_\mu$ and $A_\mu$ related by SW map are equivalent and
related to each other again by SW map. We have also explicitly computed, by using the Feynmann rules derived from the classsical action, the one-loop two-point contribution to the on-shell DeWitt action for U(1) SYM with ${\cal N}$=0,\,1,\,2 and 4 supersymmetry and found complete agreement with general result obtained by carrying out changes of variables in the path integral. These results should be
useful to guide towards a proper use of the Seiberg-Witten map.

As shown by our explicit 1-loop result, the same quadratic noncommutative IR divergences that occur in nonsupersymmetric noncommutative U(N) gauge theories formulated in terms of noncommutative fields occur in the ordinary theory obtained from the former by using the $\theta$-exact Seiberg-Witten map and that this UV/IR mixing effect --signaling a vacuum instability-- is a gauge-fixing independent characteristic of the ordinary gauge theory, in keeping with the duality statement. On the other hand, all nasty non-local noncommutative UV divergences which occur in the one-loop 1PI functional in the Feynman gauge, computed in  \cite{Horvat:2011bs,Horvat:2013rga,Horvat:2015aca,Martin:2016zon} are mere gauge artifacts since they do not occur in the one-loop two-point contribution to the on-shell DeWitt action --which is a gauge-fixing independent object-- and therefore they do not contribute to any physical quantity. Finally, the quadratic noncommutative IR diverges can be removed by considering  supersymmetric versions of the  theory, 
even though supersymmetry is not linearly realized in terms of the ordinary fields \cite{Martin:2008xa}.

One final comment regarding the validity of our all-order result: The preexistence of a self-consistent NCGFT which closes on the $\rm U(N)$ Lie algebra without Seiberg-Witten  map, admits a sound perturbative quantization by itself and, more importantly, an invertible Seiberg-Witten map. The invertibility is achievable here since the the U(N) Lie algebra generators (in the fundamental representation) form the basis of NxN complex matrices when considered as the generators of a complex linear space. The SU(N) Lie algebra, as a subspace of the U(N) Lie algebra loses this property, therefore one must require U(N) gauge symmetry for an analysis involving inversion of the SW map. There exists, however, deformed $\rm SU(N)$ gauge theories~\cite{Jurco:2000ja} possessing a noncommuativity of spacetime coordinates and having an algebra which closes on the enveloping algebra of the $\rm SU(N)$ algebra only. In that case the equivalence relation cannot be applied since we lack an intrinsic formulation of the quantum theory in terms of noncommutative fields. However, the results presented here indicate that their current definition in terms of ordinary fields by using the SW map is a sensible one, providing one uses the $\theta$-exact Seiberg-Witten map.

\section{Conclusion}

We have proven that at the quantum level the $\theta$-exact SW map provides --at least in perturbative theory with respect to the coupling constant-- a dual description, in terms of ordinary fields, of the noncommutative U(N) YM theory with or without supersymmetry. We achieve that by performing appropriate changes of variables in the path integral defining the on-shell DeWitt effective action in dimensional regularization.

There remain to be seen how the results presented here carry over to the nonpertubative regime in the coupling constant. In this regard the analysis of the nonperturbative features of ${\cal N}=2$ and $4$  supersymmetric gauge theories looks particularly interesting.




\begin{acknowledgments}
The work by C.P. Martin has been financially supported in part by the Spanish MINECO through grant FPA2014-54154-P. The work  of J.T. is conducted under the European Commission and the Croatian Ministry of Science, Education and Sports Co-Financing Agreement No. 291823 and he acknowledges project financing by the Marie Curie FP7-PEOPLE-2011-COFUND program NEWFELPRO: Grant Agreement No. 69.  J.Y. has been fully supported by Croatian Science Foundation under Project No. IP-2014-09-9582. We would like to acknowledge the COST Action MP1405  (QSPACE).
A great deal of computation was done by using MATHEMATICA 8.0~\cite{mathematica} plus the tensor algebra package xACT~\cite{xAct}. \\

$^\star$Plenary session talk delivered by Josip Trampeti\' c at the XXXVII Max Born Symposium: Wroclaw 2016, "Noncommutative and Quantum Geometries", Wroclaw, Poland, July 04-07, 2016; supported as the 2016 WG3 Meeting of COST Action MP1405.
\end{acknowledgments}

\end{document}